\documentclass[manuscript]{aastex62}

\def\rmi {\textit{r-i}\ }

 \usepackage{enumitem}
\usepackage[]{graphicx}
\usepackage{floatrow}
\usepackage{ifsym}
\usepackage{textcomp}
\usepackage{natbib}
\usepackage{csquotes}

\begin{document}

\title{First results from the rapid-response spectrophotometric characterization of Near-Earth Objects using RATIR}

\correspondingauthor{S. Navarro-Meza}
\email{snavarro@astro.unam.mx}

\author{S. Navarro-Meza}
\affil{Instituto de Astronom\'ia, Universidad Nacional Aut\'onoma de M\'exico, Ensenada B.C. 22860, M\'exico.}
\affil{Department of Physics and Astronomy, Northern Arizona University, Flagstaff, AZ 86001, USA}

\author{M. Mommert}
\affil{Lowell Observatory, Flagstaff, AZ 86001, USA}
\affil{Department of Physics and Astronomy, Northern Arizona University, Flagstaff, AZ 86001, USA}

\author{D.E. Trilling}
\affil{Department of Physics and Astronomy, Northern Arizona University, Flagstaff, AZ 86001, USA}

\author{N. Butler}
\affil{School of Earth and Space exploration, Arizona State University, Tempe, AZ 85287, USA}
\affil{Cosmology Initiative, Arizona State University, Tempe, AZ 85287, USA}

\author{M. Reyes-Ruiz}
\affil{Instituto de Astronom\'ia, Universidad Nacional Aut\'onoma de M\'exico, Ensenada B.C. 22860, M\'exico.}

\author{B. Pichardo}
\affil{Instituto de Astronom\'ia, Universidad Nacional Aut\'onoma de M\'exico, Ciudad Universitaria, D.F. 04510, M\'exico.}

\author{T. Axelrod}
\affil{Steward Observatory, University of Arizona, Tucson, AZ 85721, USA}

\author{R. Jedicke}
\affil{Institute for Astronomy, University of Hawaii at Manoa, Honolulu, HI 96822, USA}

\author{N. Moskovitz}
\affil{Lowell Observatory, Flagstaff, AZ 86001, USA}
\begin{abstract}
As part of our multi-observatory, multi-filter campaign, we present \rmi color observations of 82 Near-Earth Objects (NEOs) obtained with the RATIR instrument on the 1.5~m robotic telescope at the San Pedro Martir's National Observatory in Mexico. 
Our project is particularly focused on rapid response observations of small ($\lesssim 850$ m) NEOs.
The rapid response and the use of spectrophotometry allows us to constrain the taxonomic classification of NEOs with high efficiency. Here we present the methodology of our observations and our result, suggesting that the ratio of C-type to S-type asteroids in a size range of $\sim$30-850m is 1.1, which is in accordance with our previous results. We also find that 10$\%$ of all NEOs in our sample are neither C- nor S-type asteroids

\end{abstract}


\section{Introduction} \label{sec:intro}

Solar System minor bodies are tracers of the Solar System's formation and evolution, and hence can be used as current samples of the processes that occurred in the early days of the system and its 
formation \citep{Delse,Malhotra}. Therefore, studies with numerous samples, focused on analyzing colors, taxonomies, orbital and physical  properties of asteroids from different populations have been made  \citep[see for example][]{Ivezic, Carvano,Carry16}.\\

Near Earth  Objects (NEOs) are of particular interest for their potential to explain the disagreement between the composition of meteorite falls on Earth and the composition observed in asteroids \citep{Mommert}. Furthermore, the Chelyabinsk event in 2013 showed us that there exist NEOs with the potential to cause moderate to devastating damage to our communities \citep{chelya}. It is worth to remark that events like this can happen in any point on the globe. This event has motivated projects aimed to characterize those asteroids that could impact the Earth and that have enough energy to compromise its safety.\\

Discovery and characterization efforts aimed at NEOs have significantly increased in the last years. However, 
due to their general faintness, characterization of small NEOs lags behind. The most effective way to constrain NEO compositions is spectroscopy, which allows for the identification of both the overall continuum shape and diagnostic band features and enables their taxonomic classification. However, spectroscopy is very expensive in terms of telescope time, and is only possible with relatively bright objects, which generally are the largest objects. Only few small asteroids (with diameters smaller than 100 m) get bright enough to be observed spectroscopically \citep[e.g.][]{Moskovitz}.  Currently, \cite{Demeo} offer the most complete taxonomic classification system.\\

Photometric measurements at a few key wavelengths -\textit{spectrophotometry}- can be sufficient to estimate asteroid taxonomies (see \citealt{Mommert}, and references therein). This technique has the advantage of making faint targets accessible because the light is collected within a broad bandpass instead of being dispersed as a function of wavelength. Furthermore, spectrophotometry  can be performed with smaller telescopes than the ones necessary for spectroscopy. Here we present a combination of spectrophotometry and  \textit{rapid-response} observations, i.e., observations that are obtained shortly after the discovery of the target, when it is still relatively bright. This article constitutes the third of a series of papers oriented to taxonomically classify hundreds of small NEOs using spectrophotometry.\\

C- (organic rich) and S- (siliceous rock) taxonomic types are the dominant constituents of the distribution of large asteroids \citep{SandB,ENV}. However, the compositional distribution of small NEOs appears to be different from that of Main Belt asteroids, as well as from large NEOs. \cite{Mommert} find S and C+X\footnote{Results from \cite{Mommert} make no distinction between the C and the X taxonomic complexes. For comparison purposes the notation C+X will be used here, which stands for the set of objects corresponding to both, the C and the X complexes.} complexes to be the main components of their sample of small NEOs. Also as pointed out by them, the compositional distribution of meteorite falls does not match the observed NEO distribution, a fact that can yield information on asteroid strength. This fact needs to be studied with better statistics on the small asteroid range in order to better understand the threat to Earth from impactors. It is important to remark that according to \cite{ENV,Thomas14}, Q-type asteroids can be an important component of a magnitude biased NEO sample, due to their relatively high albedos. There are other teams interested in this same topic \cite[see for example][]{Popescu,Ieva}. \\

This study is part of a worldwide systematic survey of NEO compositions. With the use of the 3.8 m United Kingdom Infrared Telescope (UKIRT) and KMTNet-SAAO telescope, we provide detailed information on the compositional distribution of NEOs with absolute magnitudes up to \textit{H} $\sim$ 28, i.e., with a few meters in diameter. Such rapid response is generally not feasible through classical observing proposals due to heavily oversubscribed classicaly scheduled major research facilities.  Furthermore, this method allows us to classify small NEOs according to their taxonomy with a higher efficiency than current spectroscopic methods (see \citealt{Galache} for a discussion). Furthermore, photometric studies of the partial lightcurves of our objects can lead to improve the period distribution of asteroids on the smaller range  \cite[see][]{Warner}. \\

In Section \ref{sec:ratir} we describe RATIR, the multiband instrument we use. Section \ref{sec:obs} describes our rapid response approach and the planing of our observations. Section \ref{sec:analisis} addresses our data selection and analysis. In Section \ref{sec:results} we provide our results and the corresponding discussion is given in \ref{sec:disc}. Finally in Section \ref{sec:conc} we discuss the conclusions and the future outlook of this project.\\

\section{RATIR}
\label{sec:ratir}
Observations were performed with the Reionization And Transients InfraRed camera \citep[RATIR,][]{ratir}, on the 
San Pedro Martir (SPM) 1.5 m telescope at the National Mexican Astronomical Observatory (Observatorio Astron\'omico Nacional). This telescope is a Ritchey-Chr\'etien  type, operated by the Universidad Nacional Aut\'onoma de M\'exico. The instrument is equipped with two optical and two near-infrared (NIR) detectors, all of them 2048x2048 pixels. Each of these detectors corresponds to a different channel with specific filters, as shown in Table \ref{tab:detectores}. RATIR takes four images of an object in a single shot\footnote{See detailed information in RATIR's web page: \url{http://ratir.astroscu.unam.mx}} . To minimize dark current and thermal background effects, the optical detectors are water cooled, while the NIR detectors are operated in a helium-cooled cryostat.\\

RATIR was designed to study gamma ray bursts \citep[e.g.][]{nat1,nat2,nat3}, but other uses are possible as shown here (see also \citealt{Tapia, Tere,Ricci}). Observations are  executed in an automated queue mode \citep{auto} if no gamma ray burst events are ongoing. The results that we present are the first asteroid observations made with the instrument.\\

\setcounter{table}{1}
\floattable
\begin{deluxetable}{ccCl}
\tablecaption{Channels of the RATIR instrument. All detectors are 2048x2048 pixels. C0 and C1 channels hold the visual range filters (observations reported in this paper were taken with the \textit{r} and \textit{i} filters). C2 and C3 channels contain the near infrared filters; H2RG (HAWAII-2RG) are Teledyne mercury-cadmium-telluride detectors.\\
\label{tab:detectores}}
\tablecolumns{5}
\tablenum{1}
\tablewidth{0pt}
\tablehead{
\colhead{Channel} &
\colhead{Detector} &
\colhead{Field size} & \colhead{Filters} \\
\colhead{} & \colhead{} & \colhead{(arc minute square)} & \colhead{}
}
\startdata
C0 & CCD  & 5.3       & SDSS ugr and seven others  \\
C1 & CCD  & 5.3       & Fixed SDSS i        \hspace{5.5em}   \\
C2 & H2RG & 10 & Fixed WFCAM Z and Y \hspace{.5em} \\
C3 & H2RG & 10 &Fixed MKO J and H \hspace{2.1em} \\
\enddata
\end{deluxetable}

\section{Observations}
\label{sec:obs}
The observations we present here were taken during 2014 and 2015. During most part of 2015 and 2016, RATIR's channels C2 and C3 (see Table \ref{tab:detectores}) were not available due to technical problems. In this work, we analyze the set of optical-only data. Since the second half of 2016 we are obtaining observations in all four channels, which will be presented in a future publication.\\

Our rapid response approach is the key feature of this project. We trigger rapid response spectrophotometric observations of NEOs within a few days of their discovery when the objects are generally still bright enough to be observed with a 1.5 m aperture. We can observe and  characterize objects as faint as V $\sim 20$. Such rapid response is generally not feasible through classical observing programs. The first results of NEO observations made by our team are presented in \cite{Mommert} and \cite{Nic}.	\\

Potential targets are identified and uploaded into the RATIR queue on a daily basis. Accessible targets are identified among those NEOs that have been discovered within the last four weeks; this duration is partially arbitrary\footnote{After the closest approach, NEOs fade at a rate of typically 0.5 mag within one week and 5 mag within 6 weeks \citep{Galache}.}, but the method usually leads to a number of well-observable and bright potential targets. A target is considered accessible if it has a visible brightness V $\leq$ 20 and an air-mass $\leq$ 2.0, as provided by the JPL Horizons system \citep{Giorgini}, for at least the duration of the estimated RATIR integration time.
Potential targets are manually selected from the list of accessible targets, prioritizing objects with high absolute magnitudes $H_{\rm V}$\ (small sizes) and large values of \textit{H}$_{\rm V}$ -V, where V stands for the apparent magnitude of the target of the upcoming night. A high value of \textit{H}$_{\rm V}$ -V ensures that our target is observed when it is close to the Earth. RATIR queue observing scripts are automatically created for the selected targets, using the latest orbital elements of the objects of interest provided by JPL Horizons. The exposure time of each frame, as well as the total integration time in each band per visit are a function of the object's brightness. Exposure times usually ranges between 5 and 30 seconds, while the total integration time per target is usually less than 1 hour.\\

Our observations are biased in favor of bright objects. At a given distance from Earth, for objects of a given diameter, objects with higher albedo are easier to observe. For targets close to our limiting magnitude, only those with  relatively high albedos will be observed. Hence, our sample may contain a higher fraction of S and Q objects than the actual asteroid population.\\

\setcounter{table}{2}
\floattable
\begin{deluxetable}{lccccccccccc}
\tablecaption{This table presents each of the reported targets according to their number or designation, observation midtime of the observing run, and the duration of it. Also presented are, the measured color indices (solar colors have been subtracted) and corresponding uncertainties.\\
\label{tab:observations}}
\tablecolumns{12}
\tablenum{2}
\tablewidth{0pt}
\tablehead{
\colhead{Object} &
\colhead{Obs. Midtime} &
\colhead{Dur.} &
\colhead{$H_{\rm V}$} &
\colhead{\rmi} &
\colhead{error}\\
\colhead{} &
\colhead{(UT)} &
\colhead{(hr)} &
\colhead{(mag)} &
\colhead{(mag)} &
\colhead{(mag)}
}
\startdata
2014 MK6 & 2014-07-22 09:30 &   1.7 & 21.00 &  0.29 & 0.01  \\
2014 MP5 & 2014-07-21 05:25 &   1.5 & 21.80 &  0.27 & 0.02  \\
2014 OZ337 & 2014-08-04 08:13 &   0.3 & 22.50 &  0.35 & 0.02  \\
2014 QQ33 & 2014-08-25 10:20 &   1.3 & 22.10 &  0.30 & 0.02  \\
2014 TT35 & 2014-10-18 03:49 &   0.1 & 26.00 &  0.49 & 0.01  \\
2014 TZ & 2014-10-23 05:12 &   1.1 & 22.60 &  0.33 & 0.01  \\
2014 UT192 & 2014-11-10 09:03 &   0.2 & 19.60 &  0.46 & 0.03  \\
2014 UZ116 & 2014-11-03 06:41 &   1.6 & 20.90 &  0.36 & 0.04  \\
2014 WX4 & 2014-11-20 07:16 &   0.5 & 26.40 &  0.40 & 0.02  \\
2014 WZ4 & 2014-11-20 03:42 &   0.5 & 23.50 &  0.29 & 0.03  \\
2014 YE35 & 2015-01-15 08:32 &   0.5 & 20.30 &  0.38 & 0.01  \\
2014 YW34 & 2015-01-15 06:53 &   1.6 & 21.60 &  0.25 & 0.06  \\
2015 EL7 & 2015-04-05 10:19 &   0.4 & 22.70 &  0.38 & 0.03  \\
2015 EL7 & 2015-04-09 07:40 &   0.2 & 22.70 &  0.39 & 0.06  \\
2015 EL7 & 2015-04-11 07:22 &   0.1 & 22.70 &  0.36 & 0.03  \\
2015 EZ & 2015-03-15 05:56 &   0.4 & 20.30 &  0.40 & 0.01  \\
2015 FG120 & 2015-04-10 09:12 &   1.1 & 22.90 &  0.36 & 0.02  \\
2015 FG120 & 2015-04-11 09:20 &   0.2 & 22.90 &  0.29 & 0.03  \\
2015 FG120 & 2015-04-12 09:35 &   1.0 & 22.90 &  0.30 & 0.02  \\
2015 FG120 & 2015-04-13 08:07 &   0.9 & 22.90 &  0.38 & 0.02  \\
2015 FG120 & 2015-04-17 11:09 &   0.7 & 22.90 &  0.37 & 0.02  \\
2015 FG37 & 2015-04-15 10:53 &   1.1 & 21.70 &  0.41 & 0.02  \\
2015 FG37 & 2015-04-22 11:25 &   0.1 & 21.70 &  0.29 & 0.04  \\
2015 FL290 & 2015-04-09 05:11 &   1.1 & 22.20 &  0.36 & 0.02  \\
2015 FL290 & 2015-04-10 04:53 &   1.1 & 22.20 &  0.37 & 0.02  \\
2015 FL290 & 2015-04-11 04:46 &   1.2 & 22.20 &  0.39 & 0.01  \\
\enddata
\end{deluxetable}

\setcounter{table}{2}
\floattable
\begin{deluxetable}{lccccccccccc}
\tablecaption{(continued). \\
\label{}}
\title{cont}
\tablecolumns{12}
\tablenum{2}
\tablewidth{0pt}
\tablehead{
\colhead{Object} &
\colhead{Obs. Midtime} &
\colhead{Dur.} &
\colhead{$H_{\rm V}$} &
\colhead{\rmi} &
\colhead{error}\\
\colhead{} &
\colhead{(UT)} &
\colhead{(hr)} &
\colhead{(mag)} &
\colhead{(mag)} &
\colhead{(mag)}
}
\startdata
2015 FQ & 2015-03-29 05:43 &   0.9 & 22.30 &  0.40 & 0.02  \\
2015 FT118 & 2015-04-20 10:50 &   0.6 & 20.40 &  0.30 & 0.04  \\
2015 FY284 & 2015-04-02 07:13 &   0.5 & 21.60 &  0.37 & 0.05  \\
2015 GS13 & 2015-05-14 09:26 &   0.4 & 21.00 &  0.51 & 0.03  \\
2015 GS & 2015-04-15 09:34 &   1.1 & 20.60 &  0.35 & 0.02  \\
2015 GS & 2015-04-16 09:53 &   1.2 & 20.60 &  0.33 & 0.02  \\
2015 GY & 2015-04-15 05:14 &   0.6 & 21.70 &  0.41 & 0.01  \\
2015 GY & 2015-04-19 07:16 &   0.5 & 21.70 &  0.27 & 0.02  \\
2015 HA1 & 2015-04-25 09:04 &   0.9 & 21.20 &  0.46 & 0.01  \\
2015 HA1 & 2015-05-05 07:50 &   0.5 & 21.20 &  0.45 & 0.01  \\
2015 HP171 & 2015-05-12 09:32 &   0.3 & 20.10 &  0.34 & 0.02  \\
2015 HR1 & 2015-05-06 08:59 &   0.7 & 24.30 &  0.35 & 0.06  \\
2015 HR1 & 2015-05-07 09:00 &   0.4 & 24.30 &  0.34 & 0.06  \\
2015 HR1 & 2015-05-13 09:39 &   0.7 & 24.30 &  0.26 & 0.03  \\
2015 HV171 & 2015-05-08 08:55 &   0.1 & 18.10 &  0.36 & 0.01  \\
2015 HW11 & 2015-05-12 07:04 &   1.1 & 23.30 &  0.41 & 0.02  \\
2015 JQ1 & 2015-05-18 05:03 &   1.0 & 20.30 &  0.26 & 0.05  \\
2015 JQ1 & 2015-05-19 05:45 &   1.0 & 20.30 &  0.28 & 0.02  \\
2015 KL122 & 2015-06-05 09:19 &   0.5 & 22.30 &  0.47 & 0.07  \\
2015 KQ120 & 2015-05-30 10:10 &   0.1 & 26.70 &  0.43 & 0.04  \\
2015 KQ57 & 2015-05-26 05:17 &   0.0 & 22.20 &  0.40 & 0.04  \\
2015 KV18 & 2015-05-23 09:41 &   0.8 & 23.80 &  0.42 & 0.03  \\
2015 KV18 & 2015-05-24 09:28 &   0.8 & 23.80 &  0.37 & 0.04  \\
2015 KV18 & 2015-05-25 07:40 &   0.2 & 23.80 &  0.31 & 0.02  \\
2015 KV18 & 2015-05-26 07:40 &   0.1 & 23.80 &  0.41 & 0.03  \\
2015 LA2 & 2015-06-14 06:35 &   1.0 & 23.10 &  0.33 & 0.01  \\
2015 LG14 & 2015-06-22 06:22 &   1.0 & 23.20 &  0.40 & 0.03  \\
2015 LG14 & 2015-06-23 06:03 &   0.8 & 23.20 &  0.52 & 0.04  \\
2015 LG2 & 2015-06-18 07:18 &   1.0 & 20.30 &  0.40 & 0.03  \\
\enddata
\end{deluxetable}

\setcounter{table}{2}
\floattable
\begin{deluxetable}{lccccccccccc}
\tablecaption{(continued) \\
\label{}}
\title{cont}
\tablecolumns{12}
\tablenum{2}
\tablewidth{0pt}
\tablehead{
\colhead{Object} &
\colhead{Obs. Midtime} &
\colhead{Dur.} &
\colhead{$H_{\rm V}$} &
\colhead{\rmi} &
\colhead{error}\\
\colhead{} &
\colhead{(UT)} &
\colhead{(hr)} &
\colhead{(mag)} &
\colhead{(mag)} &
\colhead{(mag)}
}
\startdata
2015 LG2 & 2015-06-21 08:41 &   0.8 & 20.30 &  0.37 & 0.02  \\
2015 LJ24 & 2015-06-18 05:11 &   0.2 & 20.00 &  0.42 & 0.03  \\
2015 LJ & 2015-07-04 06:43 &   0.8 & 24.70 &  0.45 & 0.05  \\
2015 LJ & 2015-07-04 07:50 &   0.8 & 24.70 &  0.33 & 0.07  \\
2015 LJ & 2015-07-13 05:21 &   0.8 & 24.70 &  0.27 & 0.03  \\
2015 LJ & 2015-07-25 06:54 &   0.9 & 24.70 &  0.29 & 0.06  \\
2015 LQ21 & 2015-06-21 05:01 &   0.1 & 24.50 &  0.50 & 0.03  \\
2015 MC & 2015-06-20 06:18 &   0.2 & 24.10 &  0.45 & 0.02  \\
2015 MC & 2015-06-26 06:49 &   0.9 & 24.10 &  0.30 & 0.04  \\
2015 ME116 & 2015-07-14 04:57 &   0.3 & 22.30 &  0.38 & 0.06  \\
2015 ME116 & 2015-07-25 04:49 &   0.4 & 22.30 &  0.27 & 0.04  \\
2015 MQ116 & 2015-07-15 08:09 &   0.8 & 23.40 &  0.35 & 0.05  \\
2015 MS59 & 2015-07-13 10:15 &   0.9 & 21.00 &  0.25 & 0.01  \\
2015 MS59 & 2015-07-14 09:26 &   0.9 & 21.00 &  0.49 & 0.02  \\
2015 MS59 & 2015-07-15 09:25 &   1.0 & 21.00 &  0.41 & 0.02  \\
2015 MS59 & 2015-07-23 09:48 &   0.9 & 21.00 &  0.37 & 0.04  \\
2015 MU59 & 2015-07-09 10:01 &   0.8 & 20.00 &  0.31 & 0.02  \\
2015 MU59 & 2015-07-13 09:15 &   0.8 & 20.00 &  0.44 & 0.01  \\
2015 MU59 & 2015-08-04 08:44 &   0.6 & 20.00 &  0.39 & 0.01  \\
2015 MX103 & 2015-07-04 05:13 &   0.5 & 24.40 &  0.46 & 0.02  \\
2015 MX103 & 2015-07-06 06:16 &   0.6 & 24.40 &  0.38 & 0.02  \\
2015 MY53 & 2015-07-03 05:59 &   0.5 & 25.40 &  0.39 & 0.06  \\
2015 MY53 & 2015-07-03 06:49 &   0.2 & 25.40 &  0.52 & 0.03  \\
2015 NK13 & 2015-08-04 07:07 &   0.8 & 21.00 &  0.40 & 0.03  \\
2015 NK3 & 2015-08-04 05:56 &   0.8 & 21.30 &  0.26 & 0.03  \\
2015 NK3 & 2015-08-07 06:57 &   1.0 & 21.30 &  0.28 & 0.01  \\
2015 NU2 & 2015-07-21 07:10 &   0.6 & 20.90 &  0.28 & 0.04  \\
2015 NU2 & 2015-07-24 05:15 &   1.1 & 20.90 &  0.35 & 0.04  \\
\enddata
\end{deluxetable}

\setcounter{table}{2}
\floattable
\begin{deluxetable}{lccccccccccc}
\tablecaption{(continued) \\
\label{}}
\title{cont}
\tablecolumns{12}
\tablenum{2}
\tablewidth{0pt}
\tablehead{
\colhead{Object} &
\colhead{Obs. Midtime} &
\colhead{Dur.} &
\colhead{$H_{\rm V}$} &
\colhead{\rmi} &
\colhead{error}\\
\colhead{} &
\colhead{(UT)} &
\colhead{(hr)} &
\colhead{(mag)} &
\colhead{(mag)} &
\colhead{(mag)}
}
\startdata
2015 OF26 & 2015-08-07 05:37 &   0.3 & 21.60 &  0.32 & 0.02  \\
2015 OM21 & 2015-07-24 09:55 &   0.7 & 22.50 &  0.38 & 0.03  \\
2015 OM21 & 2015-08-06 08:00 &   0.9 & 22.50 &  0.45 & 0.03  \\
2015 OM21 & 2015-08-07 08:05 &   1.0 & 22.50 &  0.48 & 0.02  \\
2015 PA229 & 2015-08-21 09:39 &   1.0 & 21.40 &  0.41 & 0.02  \\
2015 PA229 & 2015-09-05 08:55 &   1.0 & 21.40 &  0.39 & 0.05  \\
2015 PQ56 & 2015-09-03 07:59 &   0.9 & 22.60 &  0.46 & 0.04  \\
2015 PQ & 2015-09-07 07:52 &   0.7 & 22.70 &  0.48 & 0.01  \\
2015 QB & 2015-08-21 08:32 &   1.0 & 24.20 &  0.39 & 0.01  \\
2015 QG & 2015-08-22 05:17 &   0.3 & 23.80 &  0.33 & 0.02  \\
2015 QM3 & 2015-08-23 04:45 &   0.4 & 20.40 &  0.28 & 0.05  \\
2015 QN3 & 2015-08-23 04:20 &   0.4 & 19.50 &  0.28 & 0.01  \\
2015 QN3 & 2015-08-24 04:56 &   0.4 & 19.50 &  0.40 & 0.01  \\
2015 QO3 & 2015-08-24 07:05 &   0.6 & 19.40 &  0.38 & 0.01  \\
2015 RH36 & 2015-09-18 09:57 &   0.5 & 23.60 &  0.37 & 0.06  \\
2015 RO36 & 2015-09-18 05:36 &   0.3 & 22.90 &  0.24 & 0.03  \\
2015 RQ36 & 2015-09-16 07:25 &   1.0 & 24.50 &  0.36 & 0.01  \\
2015 RQ36 & 2015-09-19 09:05 &   0.6 & 24.50 &  0.35 & 0.02  \\
2015 SO2 & 2015-09-26 09:01 &   0.3 & 23.90 &  0.37 & 0.03  \\
2015 SO2 & 2015-09-27 09:33 &   0.3 & 23.90 &  0.29 & 0.03  \\
2015 SO2 & 2015-09-28 10:57 &   0.3 & 23.90 &  0.50 & 0.03  \\
2015 SO2 & 2015-10-02 11:21 &   0.4 & 23.90 &  0.32 & 0.03  \\
2015 SV2 & 2015-09-29 05:31 &   0.9 & 20.80 &  0.33 & 0.03  \\
2015 SY & 2015-10-01 06:35 &   0.7 & 23.30 &  0.44 & 0.02  \\
2015 SY & 2015-10-02 06:09 &   0.8 & 23.30 &  0.33 & 0.03  \\
2015 SZ & 2015-10-02 05:22 &   0.4 & 23.50 &  0.36 & 0.01  \\
2015 TE & 2015-10-08 04:26 &   0.5 & 22.50 &  0.43 & 0.02  \\
\enddata
\end{deluxetable}

\setcounter{table}{2}
\floattable
\begin{deluxetable}{lccccccccccc}
\tablecaption{(continued) \\
\label{}}
\title{cont}
\tablecolumns{12}
\tablenum{2}
\tablewidth{0pt}
\tablehead{
\colhead{Object} &
\colhead{Obs. Midtime} &
\colhead{Dur.} &
\colhead{$H_{\rm V}$} &
\colhead{\rmi} &
\colhead{error}\\
\colhead{} &
\colhead{(UT)} &
\colhead{(hr)} &
\colhead{(mag)} &
\colhead{(mag)} &
\colhead{(mag)}
}
\startdata
2015 TF & 2015-10-10 05:15 &   0.6 & 22.20 &  0.39 & 0.01  \\
2015 TW178 & 2015-10-26 04:18 &   0.9 & 21.20 &  0.28 & 0.04  \\
2015 TY144 & 2015-10-26 10:14 &   0.5 & 21.30 &  0.48 & 0.05  \\
2015 TY178 & 2015-11-06 06:52 &   0.7 & 21.80 &  0.32 & 0.04  \\
2015 UJ51 & 2015-10-27 07:41 &   0.7 & 21.40 &  0.38 & 0.03  \\
2015 US51 & 2015-10-28 04:54 &   0.6 & 22.40 &  0.44 & 0.02  \\
2015 US51 & 2015-11-01 05:02 &   0.8 & 22.40 &  0.43 & 0.01  \\
2015 US51 & 2015-11-02 04:35 &   0.8 & 22.40 &  0.44 & 0.01  \\
2015 UT52 & 2015-11-05 07:40 &   0.8 & 20.90 &  0.45 & 0.03  \\
2015 UT52 & 2015-11-11 10:56 &   0.6 & 20.90 &  0.40 & 0.04  \\
2015 VJ2 & 2015-11-10 09:36 &   0.4 & 19.60 &  0.44 & 0.02  \\
2015 VJ2 & 2015-11-18 10:50 &   0.2 & 19.60 &  0.37 & 0.03  \\
2015 VJ2 & 2015-11-19 10:06 &   0.6 & 19.60 &  0.26 & 0.01  \\
2015 VO66 & 2015-11-14 10:06 &   0.4 & 20.60 &  0.25 & 0.03  \\
2015 VO66 & 2015-11-19 08:20 &   0.5 & 20.60 &  0.28 & 0.01  \\
2015 VZ2 & 2015-11-19 04:15 &   0.9 & 22.70 &  0.46 & 0.04  \\
2014 OT338 & 2014-08-17 11:10 &   1.4 & 21.40 &  0.45 & 0.02  \\
2014 TX32 & 2014-10-16 05:35 &   1.1 & 20.20 &  0.42 & 0.01  \\
2015 DE176 & 2015-02-28 06:18 &   1.0 & 19.70 &  0.35 & 0.03  \\
2015 JV & 2015-05-19 07:35 &   0.8 & 21.50 &  0.33 & 0.01  \\
2015 KJ19 & 2015-05-24 04:53 &   0.6 & 22.50 &  0.28 & 0.07  \\
\enddata
\end{deluxetable}

\section{Data Analysis}\label{sec:analisis}

 \subsection{Taxonomic classification}\label{sec:tax}
We use the Bus-DeMeo classification scheme \citep{Demeo} to classify our sample. This is a widely used taxonomic scheme that combines the visible and near infrared ranges, covering from 0.45 to $2.45 \ \mu$m. The taxonomy includes 24 classes, most of which correspond to the C-, S- and X-complexes that include the majority of the known asteroids \citep[see][and Section \ref{sec:intro}]{Demeo}. For this reason, we considered these 3 complexes in our analysis, as well as the Q-type, which, as described by \cite{ENV,Thomas14}, can be an important component of a magnitude biased NEO sample like ours.  Other taxonomic types were not considered, as 	they are not expected to be a significant part of the distribution (perhaps up to $20\%$: \cite{Mommert,Nic,Perna,Lin}) and due to the simplicity of our model. We revisit this assumption below.\\

\begin{figure}
\centering
{\caption{All \rmi indices considered in this work according to the Bus-DeMeo taxonomies.  Orange lines are the subtypes that are most distant from the C- and S-type respectively, thus defining the limits of these complexes . Notice that the C- and S-subtype are in the middle of their complexes.}\label{fig:types}}
{\includegraphics[clip=true,trim=21 0 0 140]{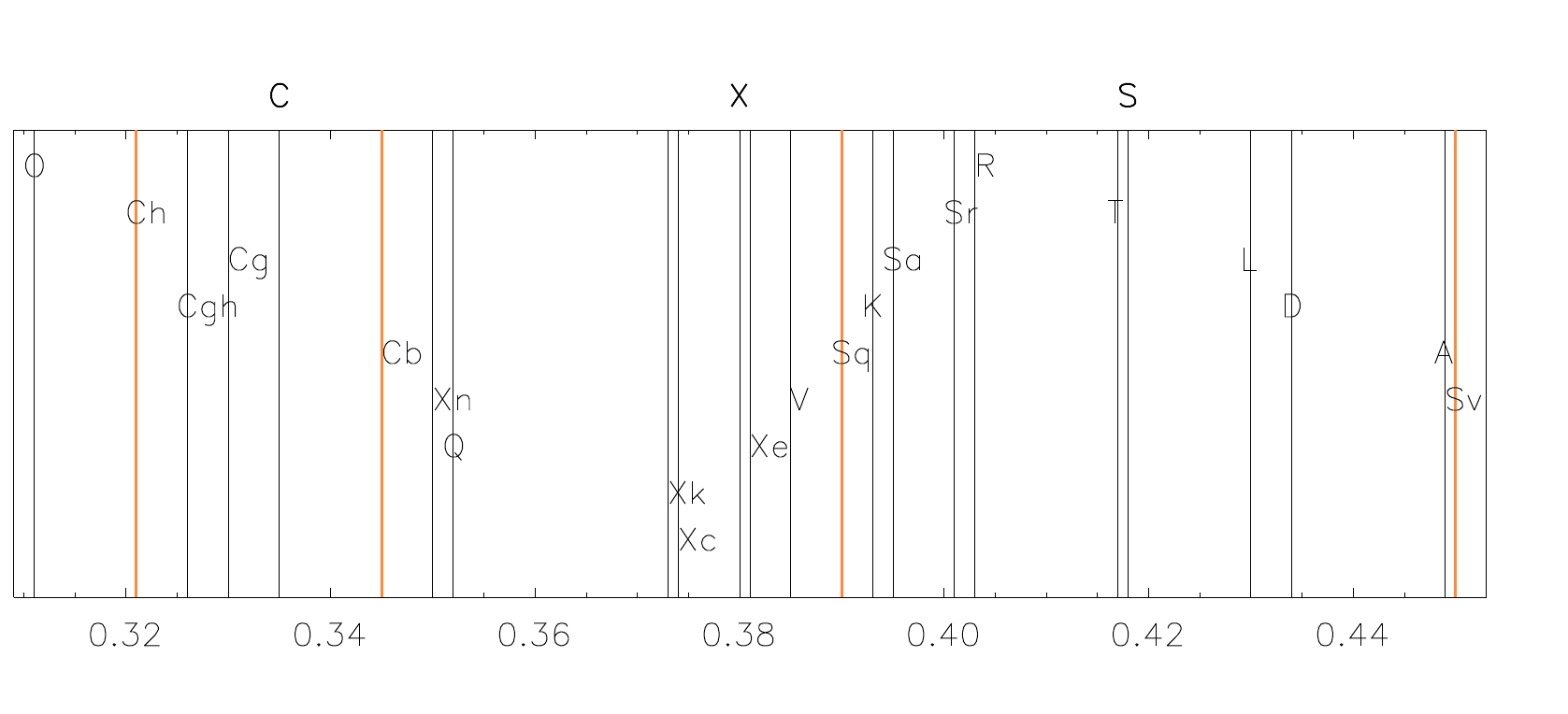}}
\end{figure}

We obtain the characteristic color of each taxonomic type from a sample of measured asteroid spectra\footnote{http://smass.mit.edu/minus.html}. For each object from the sample, its reflectance spectrum is convolved with the spectral response of each RATIR's filter and the solar spectrum. Details of the process can be found at \cite{Mommert}, Section 3. Figure \ref{fig:types} shows these color indices in the \rmi color.\\

\subsection{Photometry} \label{sec:pipe}
 Image reduction and photometry is carried out using a pipeline developed for GRB observations \citep[see, e.g.][]{Becerra, LJ}. Briefly, the pipeline reduces, sky-subtracts, and aligns input frames. These frames are then stacked into a sky image. The stellar PSF is determined and fit using custom python scripts to determine the photometric zero point in comparison with SDSS, 2MASS, and/or USNO photometric catalogs. After finishing the GRB pipeline reduction, we create a source mask using the sky image. By then coadding the frames in the moving frame of the target, keeping track of the exposure per pixel, the non-moving sources are removed and we retain only the signal from the target.  The PSF determined from the sky image is then fit to the moving-target image and the zero point from the sky image is applied to normalize the photometry.\\
 
 The magnitude of an object as a function of the observing time is generated by dyadically combining the masked frames, selecting a sufficiently long time interval for each photometric epoch as to adequately fill in masked pixels prior to PSF fitting.  In principle, single frame photometry is possible because we propagate the exposure pixel by pixel; however the accuracy can depend strongly on the stability of the PSF.  \\

 Therefore, the pipeline yields photometry on the original image, on a set of stacked images, and on the overall visit's stacked frame. The stacking creates new images with different virtual exposure times which are integer multiples of the real exposure time. The result of this procedure is available in data tables and through a graphical display in a website. In our analysis we use the photometry measured in individual \textit{r} band and \textit{i} band images. With this information we measure the \rmi color index. The Solar \rmi was subtracted from our measurements in order to make them compatible with the synthetized colors (see Section \ref{sec:tax}).\\

\subsection{Outlier rejection}

\begin{figure}
  \begin{center}
   \includegraphics[scale=.7,clip=true,trim = 10 5 8 10]{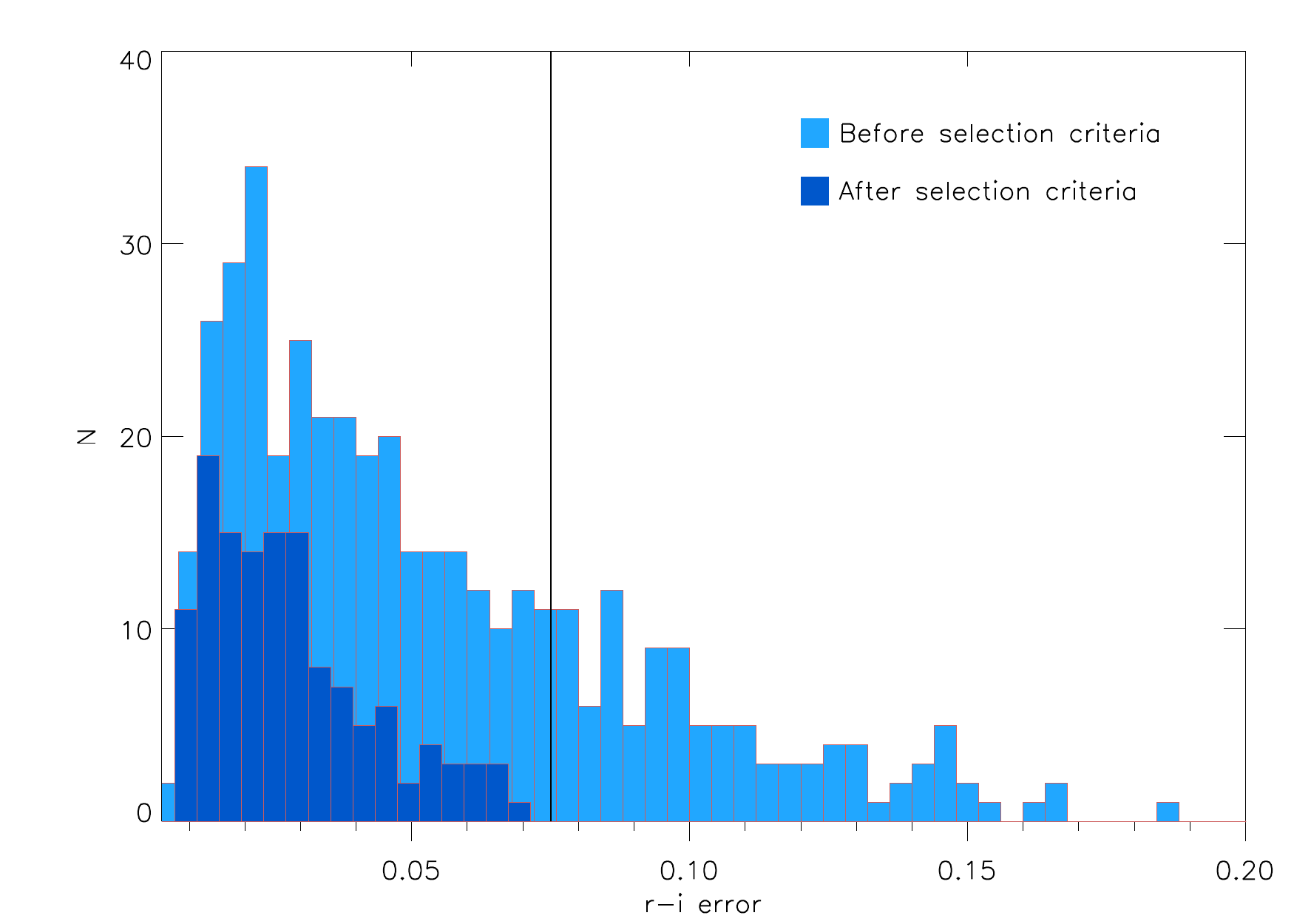}
  \end{center}
  \caption{Error distribution of the color from our sample. The vertical line shows the upper error limit set for the sample. See Section \ref{ok} for details on the selection criteria used.\\
  }
  \label{fig:errors}
 \end{figure}

In order to reject photometric outliers we performed a $10-\sigma$ clipping on the \rmi index for each visit: a weighted mean of the \rmi index was taken, then any measurement further than $10\sigma$ from the mean was rejected and the weighted mean calculated again. This process was carried out 3 times. Each time the photometric errors from the individual measurements were propagated to obtain the weights \citep{Taylor}. Therefore, the corresponding error on the \rmi index from a visit is:

 \begin{equation}
  \varsigma = \frac{1}{\sqrt{\Sigma w_{\rm k}}},
  \label{werror}
 \end{equation}

 \noindent where

  \begin{equation}
   w_{\rm k} = \frac{1}{e_{\rm k}^2} \equiv \frac{1}{e_r^2 + e_i^2}  
 \end{equation}

 \noindent where $e_{\rm k}$ is associated with the \rmi from each of the non-rejected data points of that visit and $e_r$  and $e_i$ are the photometric errors from the \textit{r} and \textit{i} band measurements. The values of $\varsigma$ are plotted in Figure \ref{fig:errors}.\\

\vspace{1em}

 \subsection{Selection of the best observations}{\label{ok}}
 We only consider measurements of those objects that passed through all of our selection criteria, the first of which is a clean visit-stacked image: a well defined source and a successful removal of the not moving sources (see Section \ref{sec:pipe} for details on the photometry). Note that we use the visit stacked image only to check the quality of the photometry and of the observation itself, e.g., with respect to background sources confusion. Also, a limit on the color index's error due to photometric uncertainty must be set. The difference in color index between the C- and S-type asteroids is 0.084, hence it is convenient that we only consider the objects that have an error lower than this threshold. Based on the discussion on Section \ref{sec:gaussian}, we decide to use 0.075 as an upper error limit on the color determination. We require a minimum of 4 measurements per visit.\\

 The outcome of this selection process is 82 different objects observed in 131 visits.\\
 
 \subsection{Probability density }\label{sec:pdf}
\begin{figure}
\centering
 \includegraphics[clip=true,trim=5 0 10 0,scale=0.7]{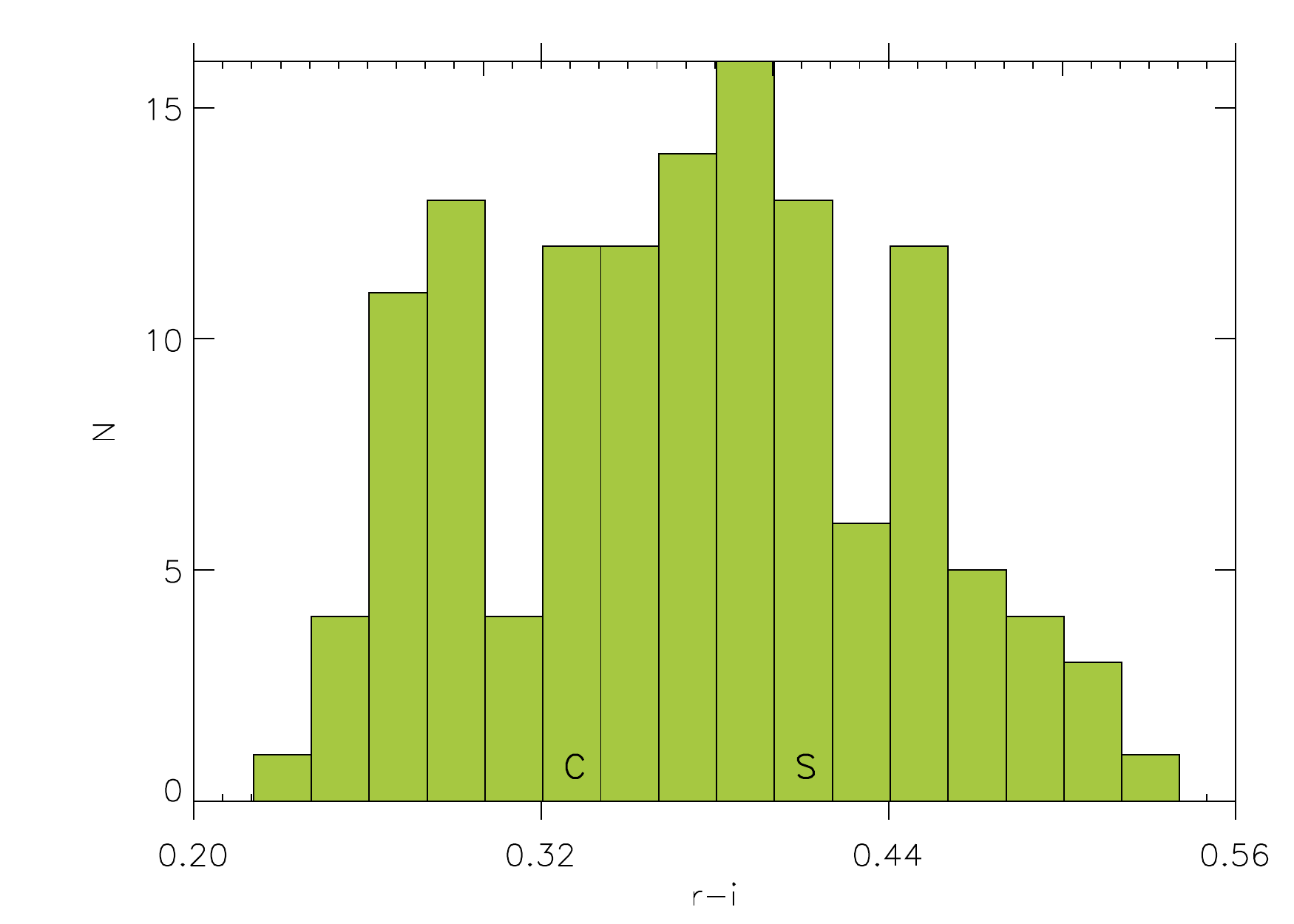}
 \caption{Color distribution of our sample. Color indices were obtained after performing the rejection process described in the text. The main subtypes from the C- and S-complex are shown (see Figure \ref{fig:types}).}
 \label{fig:histo}
\end{figure}

\begin{figure}
\centering
 \includegraphics[clip=true,trim=35 0 10 0,scale=0.7]{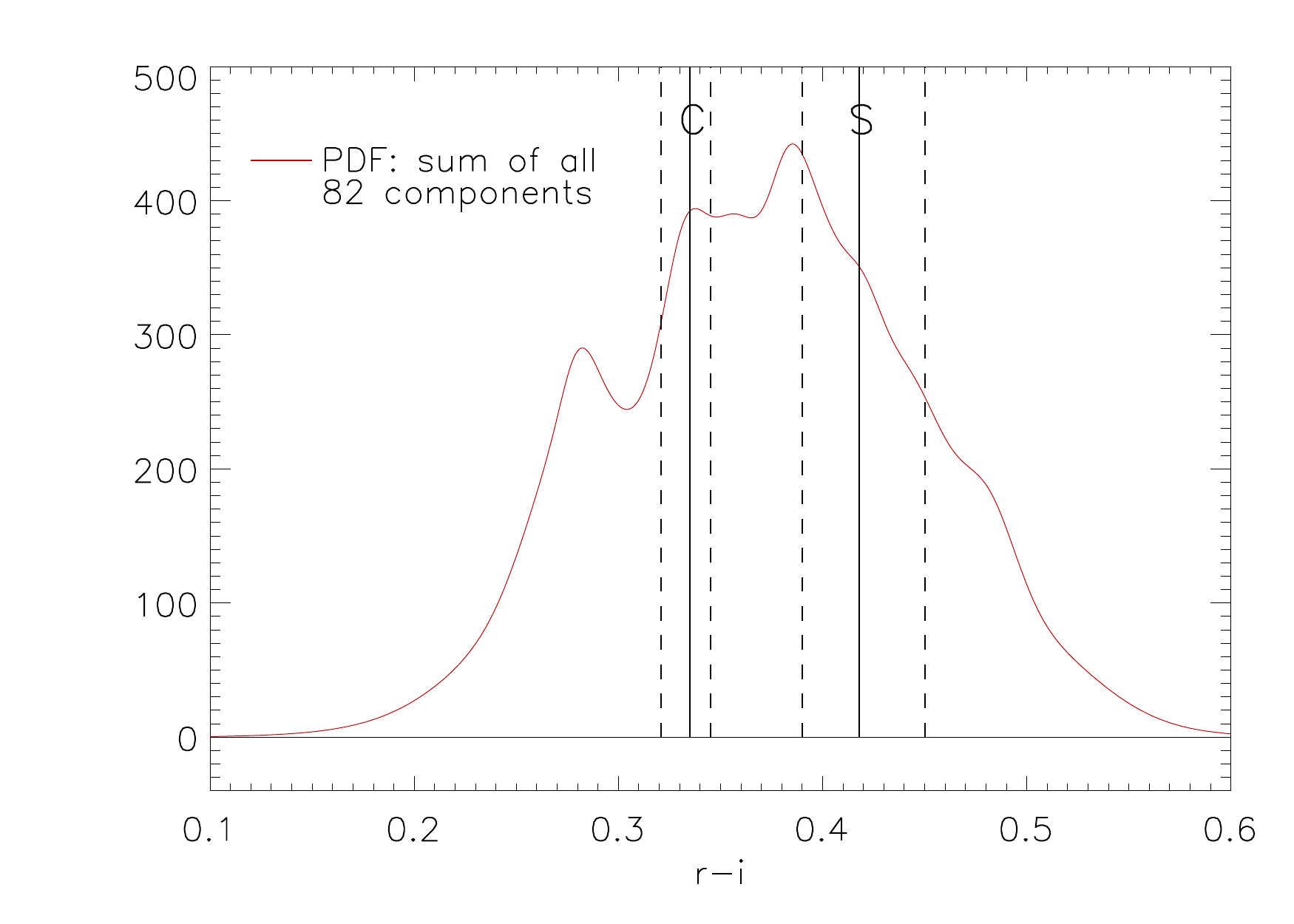}
 \caption{Probability Density Function of the color in our sample. Every visit is considered as a Gaussian centered in the color index obtained. Vertical lines represent the limits of the C and S complexes. See text for full explanation.}
\label{fig:dist}
\end{figure}

 After the selection process described in the previous section, we have one \rmi index and its associated error for each visit in our clean sample. These indices are shown in Figure \ref{fig:histo}.  In order to analyze the taxonomic distribution of our sample, we model it based on the known asteroid colors. We consider every count in Figure \ref{fig:histo}  as a normalized Gaussian centered at the \rmi value of that object with the width of the Gaussian equal to the error on the \rmi index (the height is therefore a free parameter). For objects observed more than once, the Gaussian's normalization factor is divided by the number of visits each object has. That way an object observed in more than  one visit will be represented by different Gaussians all of which add up an area of unity. The Gaussians corresponding to all objects were added up to obtain the Probability Density Function (PDF), shown in Figure \ref{fig:dist}.\\
  
 \subsection{Monte Carlo simulation}
 \label{sec:mc}
  In order to measure the compositional distribution from the PDF, we conducted a Monte Carlo (MC) simulation, which consists of creating $10^7$ samples of synthetic asteroids, each sample with the same number of objects as our data sample but with different taxonomic distributions. In each run the S-, C-, X- complexes and the Q-type are considered, and the number of objects in each taxonomy is set with a pseudo-random number generator.
  We call the resulting color index distribution from each run the Random Probability Density Function, RPDF, to distinguish it from the PDF from our data. The process to generate each distribution is the same. The details of our MC simulation are as follows.\\

82 synthetic objects were used in each MC run in order to directly compare the RPDF with the PDF. From  \cite{Mommert} and \cite{Nic}, we expect the C- and S-type asteroids to be the main components of our sample, and to find a $C/S$ ratio of $\sim$1. However, we take into account the possibility that our sample is composed of any combination of the taxonomic types considered. This is achieved by using a pseudo-random number generator under a uniform distribution with equal weights for the 3 complexes and the Q-type. The process to obtain the composition on each of the runs is as follows.\\

The order in which the number of elements is set for each complex and the Q type is randomly sorted. Each of them is labeled as $n^{\rm a}, \ n^{\rm b},\  n^{\rm c}$ or $\ n^{\rm d}$. The assignation of number of elements is always in alphabetical order, however the correspondence between the types and $n^{\rm a-d}$ is given by the pseudo random number generator. Then $n^{\rm a}$ can be assigned with any number between 0 and 82, the availability for $n^{\rm b}$ is 82-$n^{\rm a}$, and similar for $n^{\rm c}$ and $n^{\rm d}$.\\ 

The main subtypes of the complexes, respectively C-type, S-type and X-type, are the most likely to be present in our sample \citep{aiv}. Therefore, in each of the random generated samples, half of the elements assigned to each complex are given to the main subtype, while the other half is uniformly distributed among the other subtypes (see Figure \ref{fig:types} for the \rmi index of each of the members of these complexes). The number of elements in the second half will likely not be an integer multiple of the number of subtypes. For example, if 36 elements are assigned to the S-complex, 18 will correspond to the main type, the S-type. The remaining 18 will correspond to the other 4 subtypes of the complex, but 18 over 4 is not an integer. The procedure is therefore as follows: if the number of elements assigned to the 4 subtypes are $\mathbf{ n_{\rm S1}, \ n_{\rm S2},\ n_{\rm S3},\ n_{\rm S4}}$, then: \newline $\mathbf{n}_{\rm S1}=$ round(18/4) $=$ 5; 18-5 = 13 elements available for the 3 other subtypes, \newline $\mathbf{n}_{\rm S2}=$ round(13/3) $=$ 4; 13-4 $=$ 9, 
\newline $\mathbf{n}_{\rm S3} =$  round(9/2) \ \ $=$ 5; 9-5 $=$ 4, \newline $\mathbf{n}_{\rm S4} =$  4.\\

\begin{figure}
\centering
 \includegraphics[clip=true,trim=0 120 150 0,scale=0.6]{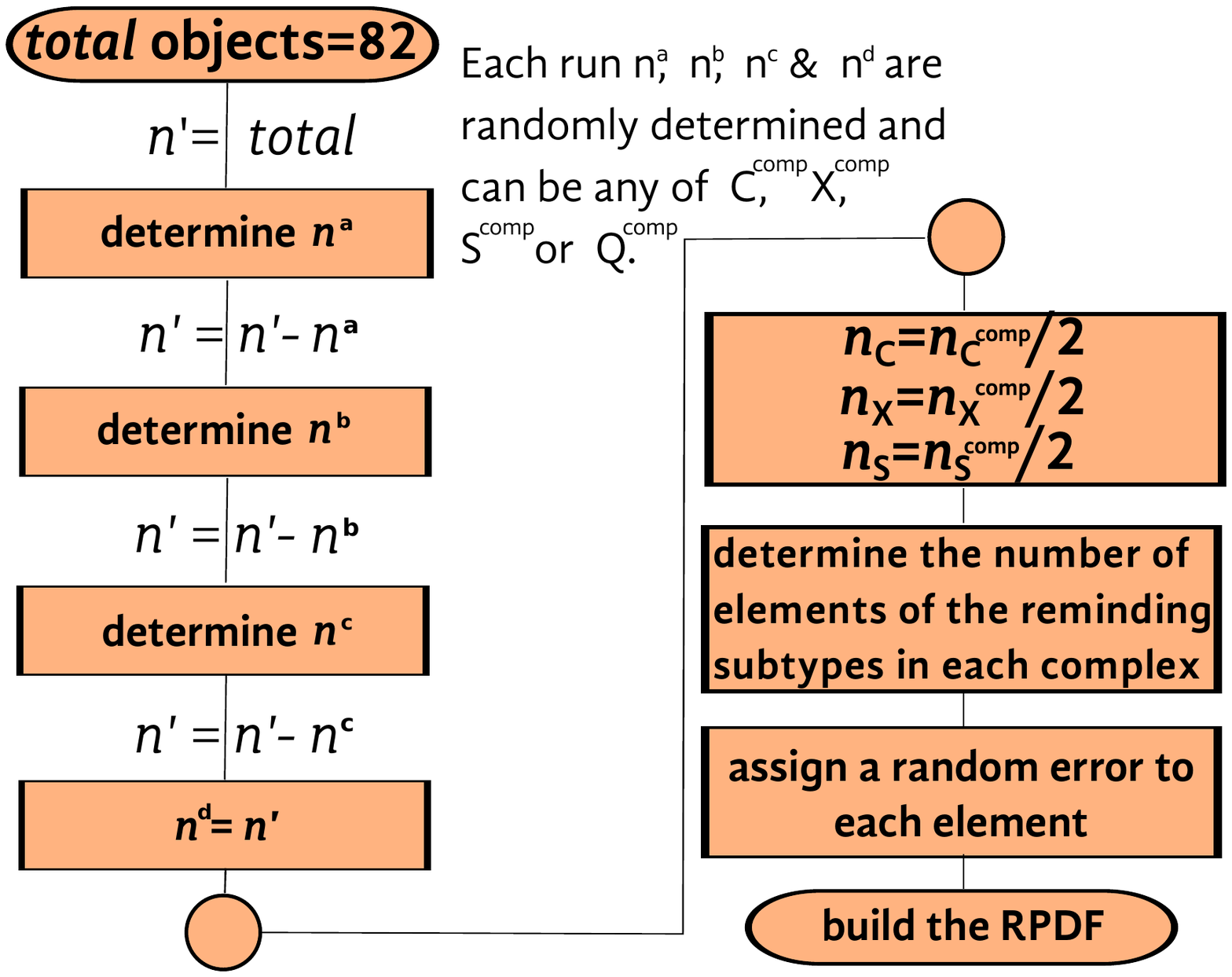}
 \caption{Main processes in the Monte Carlo simulation.
Particularly the procedure for setting the number
of objects of each type  in the synthetic sample. 
Every box including the word \enquote{determine} involves
a random process.  \textit{n} is the number of objects of
 a certain type or complex in a single run; the
 subindex indicates the referred taxonomy.
 Since there are C-, S- and X- complexes, as well as a
 C-, S- and X-types, a super index \enquote{comp} indicates
 when the variable is associated to the complex. 
\textit{n}' is the number of objects available for the unassigned
taxonomies at a certain point.}
\label{fig:mc}
\end{figure}

In each run, the correspondence between the $n_{S1-4}$ and the 4 subtypes is randomly sorted, so that over the $10^7$ samples generated, none of the 4 subtypes is favored. The same criteria apply for the C- and the X-complex. The main processes of the simulation are represented in Figure \ref{fig:mc}.\\

 After applying all the selection criteria (see Section \ref{ok}), some of the objects that were observed during different visits showed a different \rmi index. This fact was not considered in the building of the RPDF. In the simulation, the elements that are not members of the C- or the S-complex are defined as \enquote{pollution}.\\

Once the number of elements of each type in a single run is determined, it is necessary to add an error to them in order to emulate the photometric uncertainty. In order to take this into account, we fitted a Gaussian function to the distribution from Figure \ref{fig:errors} and created a random distribution under that function. 82 errors from that distribution were assigned to the \rmi indices of each of the determined types. This completes our generated sample. 82 elements were randomly selected following a distribution based on current NEOs observations. Each of the elements has an associated error, based on the error distribution of our observations. Having these, the RPDF was built up. After creating the $10^7$ samples, the PDF was compared to each of the $10^7$ RPDFs. \\

 \section{Results}
  \label{sec:results}
\begin{figure}
\centering
\caption{$\chi_{\rm r}^2$ from the $10^7$ MC simulations. Vertical line shows the domain of the first percentile of simulations in terms of the $\chi_{\rm r}^2$. \label{fig:chi}}
{\includegraphics[clip=true,trim=0 0 0 0] {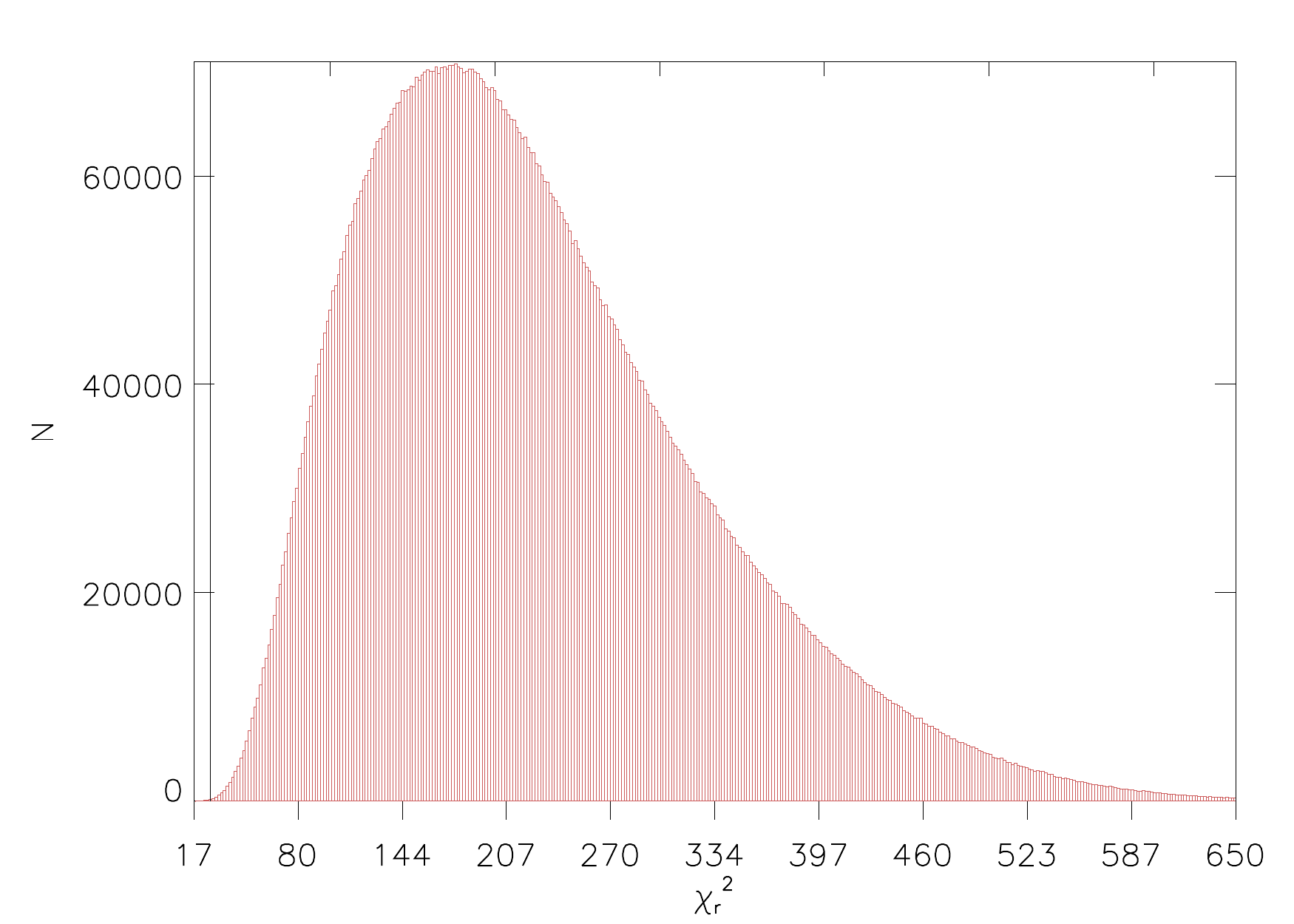}}
\end{figure}

\begin{figure}
\centering
\caption{C/S distribution from the 1st percentile of our MC simulations in terms of $\chi_{\rm r}^2$. Note: the element in the far right is the 66th best fit, therefore not considered an issue but still considered in the estimation of the mean and standard deviation. \label{fig:alfa}}
{\includegraphics[clip=true,trim=0 0 0 0] {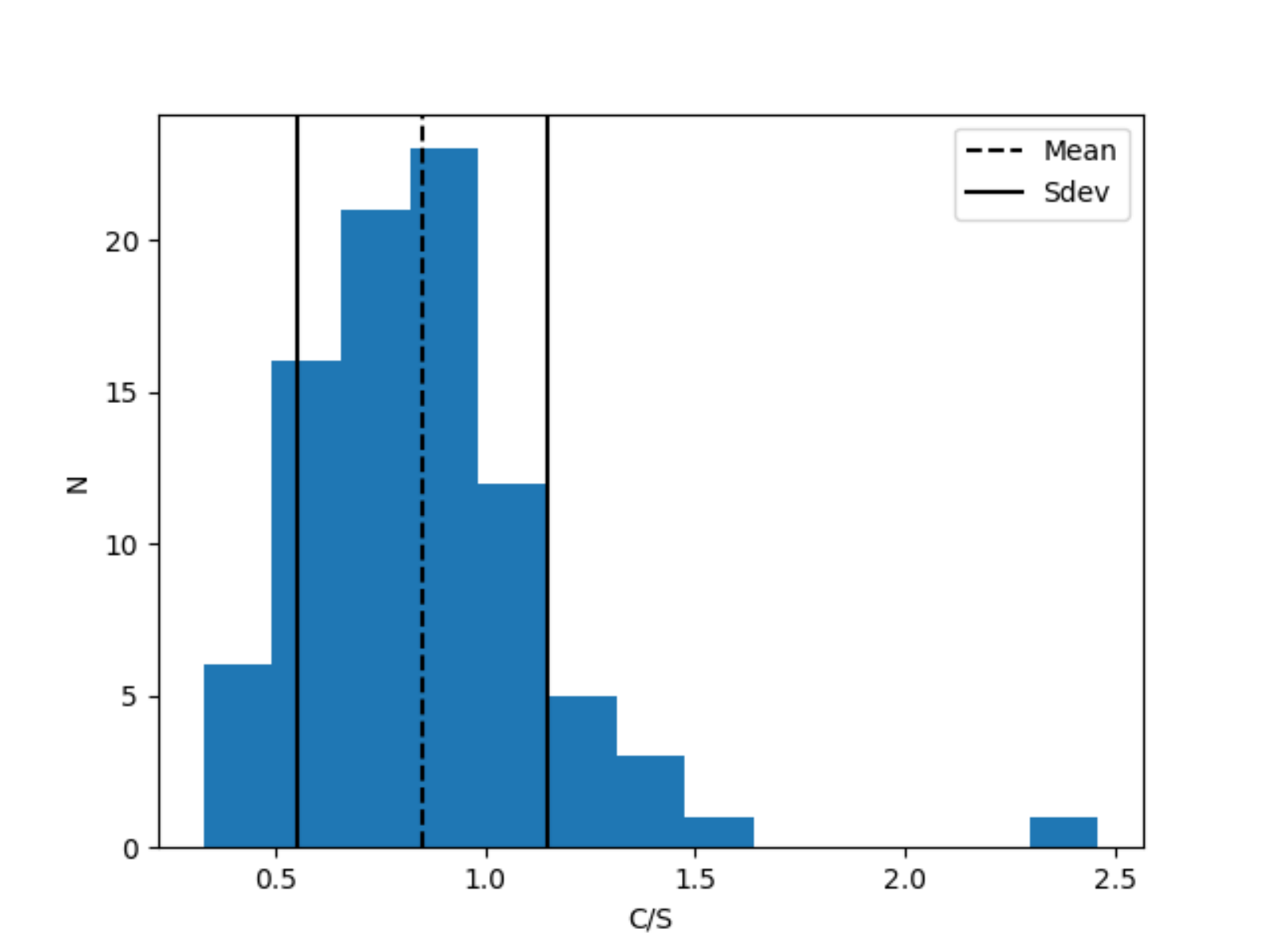}}
\end{figure}

  In order to extract the results from the simulations, we calculated the reduced $\chi^2$ ($\chi_{\rm r}^2$) between the RPDF and the PDF. Since we are allowing the simulation to create any possible combination of the taxonomies considered, the $\chi_{\rm r}^2$ range is wide as can be seen in Figure \ref{fig:chi}. We took the first percentile of simulations in terms of the $\chi_{\rm r}^2$. This is a set of 88 simulations ($\sim 10^{-5}$ from the total). The difference in $\chi_{\rm r}^2$ between the best and the second to the best case is minimal. Both present 36 elements of the S-type, the first one suggest 38 C-types, while the second 39. The main difference is in the number of X- and Q-type objects. To analyze the behavior of the best cases  we calculated the distribution of the ratio $C/S$, which is shown in Figure \ref{fig:alfa}. Our best case corresponds to a value of 1.06 in this space. Notice that value is well within 1 standard deviation from the mean, but not in the bin where mean of the distribution is. We ascribe this to the fact the histogram is skewed to the left, which is discussed in the next Section. Table \ref{tab:results} shows the percent compositions of the 4 taxonomic types considered according to our best case. The errors correspond to the standard deviation of the individual distribution of each class within the 1st percentile. Although we present results on the 3 complexes and the Q-type, the scope of this analysis is restricted to suggest a $C/S$ ratio. All of our targets have a subkilometer diameter. We don't report subdivisions in size since there was not a clear trend on the results by doing so.\\ 
  
\floattable
\begin{deluxetable}{cc}
\tablecaption{Compositional fractions found in our Monte Carlo simulation. \label{tab:results}}
\tablecolumns{2}
\tablenum{4}
\tablewidth{0pt}
\tablehead{
\colhead{Taxonomic type} &
\colhead{Percentage}\\
}
\startdata
C& 46 $\pm 9$ \\
S& 44 $\pm 8$ \\
X& 8  $\pm 9$ \\
Q& 2  $\pm 8$ \\
\enddata
\end{deluxetable}

 \begin{figure}
\centering
 \includegraphics[clip=true,trim=35 0 10 0,scale=0.7]{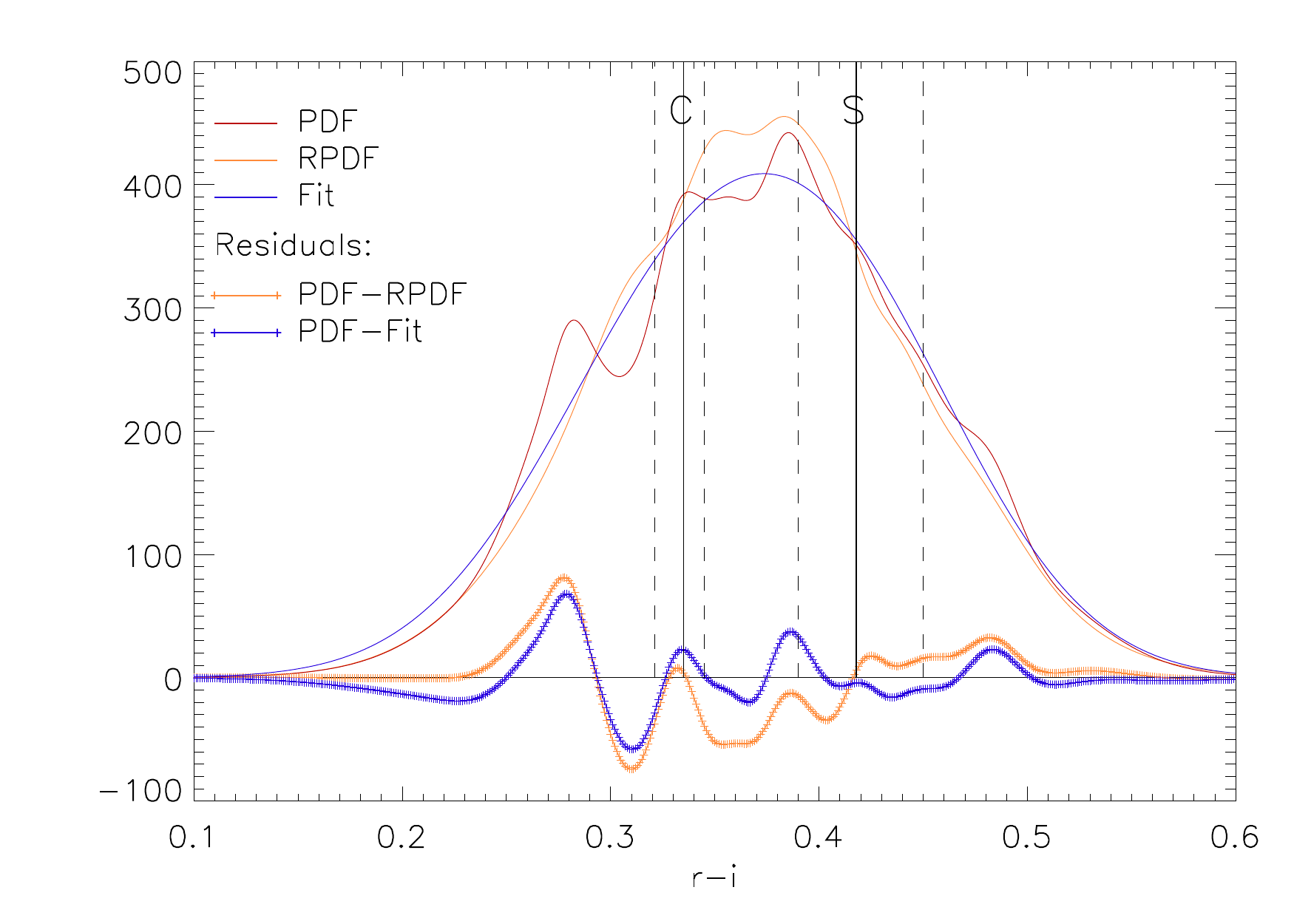}
 \caption{Probability Density Function (red) of the color in our sample. Equivalent to Figure \ref{fig:dist}, this time the best result from our MC simulation is overplotted (Random Probability Density Function, orange). A Gaussian fit to the PDF is also overplotted (blue). Both, the RPDF and the Fit make a good match with the data, as can be seen on the residuals (dashed orange and blue). See Sections \ref{sec:results} and \ref{sec:gaussian} for details.
}
\label{fig:rpdf}
\end{figure}

\section{Discussion}
 \label{sec:disc}

\subsection{Limitations and comparison with previous studies}
The strongest bias in our sample is the one presented by albedo. Our rapid-response approach is based on optically discovered objects, hence the sample targets are more likely to have moderate to high surface albedos. This can lead to an over-estimated fraction of S objects, which means the real fraction of $\rm C/S$ 
 can be higher than the estimated here. Hence, our sample is biased, and although debiasing of NEOs had been carried out \citep{SandB,mary} we will address it in a future work. \\

Our full sample contains observations in up to 6 different bands in the optical and near infrared range. However, the results presented here correspond to our control sample using only the \rmi color. Hence, one the main goals of this article is to describe the methodology we are using for our observations. Our future work will include data in all bandpasses.\\

The distribution in Figure \ref{fig:alfa} is skewed to the left due to the feature at \rmi $\sim 0.26 - 0.30$ in Figures \ref{fig:histo} and \ref{fig:dist}. It is likely due to a systematic error in the observations causing a bluer  color. The analysis of the complete sample will allow us to test this idea.\\

Our full sample is one of the largest for small NEOs. The analysis of it as well as the work from other teams is needed for a comprehensive classification of this kind of objects. (As a comparison, \cite{Ivezic} made a study of the Main Belt including $\sim$ 13,000 objects). So far, the results from other teams on a similar size range to ours (subkilometer) are compared next. The fraction of S-type we find is in agreement within the error bars of the other studies. \cite{Ieva}, from the analysis 67 NEOs found $\sim 61\%$ of S-types. \cite{Lin}, presents 51 subkilometer NEOs. With this sample they found an S fraction of $\sim 33\%$. \cite{Perna}, for a sample of 146 objects, found an S fraction of $\sim 40\%$. Our team, using different telescopes and samples than the one presented in this paper, found $\sim 40\%$ with 40 NEOs in \cite{Mommert}, while \cite{Nic}, with a sample of 45 objects, obtained an S fraction of $\sim 43\%$. Using a sample of 252 objects \cite{SandB} found an S-type fraction of $22\%$. This number is a reference for the distribution of NEOs, but is not directly comparable with our results since they performed bias correction, their sample includes Mars Crossing asteroids, and includes objects up to the 10 km scale. \cite{Binzel} (in press) uses a sample of 1040 objects, with a median of 0.7 km and finds an S fraction of $\sim 50\%$. The C-type fraction found in these same papers varies and can be as low as $10\%$ vs the $\sim 46\%$ we suggest.\\

  \subsection{Gaussian Fit}
\label{sec:gaussian}
With the purpose of getting a result independent from the MC simulation, we performed a Gaussian fit on the PDF, for which, we considered two components with fixed mean: one centered on the color index of C-type asteroids (0.335), and the other one centered on the color index of S-type asteroids (0.418).\\

The result from our fit is overplotted to the PDF and RPDF in Figure \ref{fig:rpdf}. The two components of the fit are equivalent to 48$\%$ of S-type objects and 52$\%$ of C-type objects yielding $C/S=1.07$, which is the same value than in the MC result. The residual (dashed blue in the figure) does not show a systematic
behavior, and it is similar to the residual of subtracting the RPDF from the PDF
(dashed yellow curve in the figure), so we assume this is primarily noise.\\

Fits considering other taxonomic types were made, yielding a larger residual. Because of this observation we decided only to consider the C- and S-type in the fitting process.\\

\begin{figure}
\centering
 \includegraphics[clip=true,trim=10 0 10 10,scale=0.7]{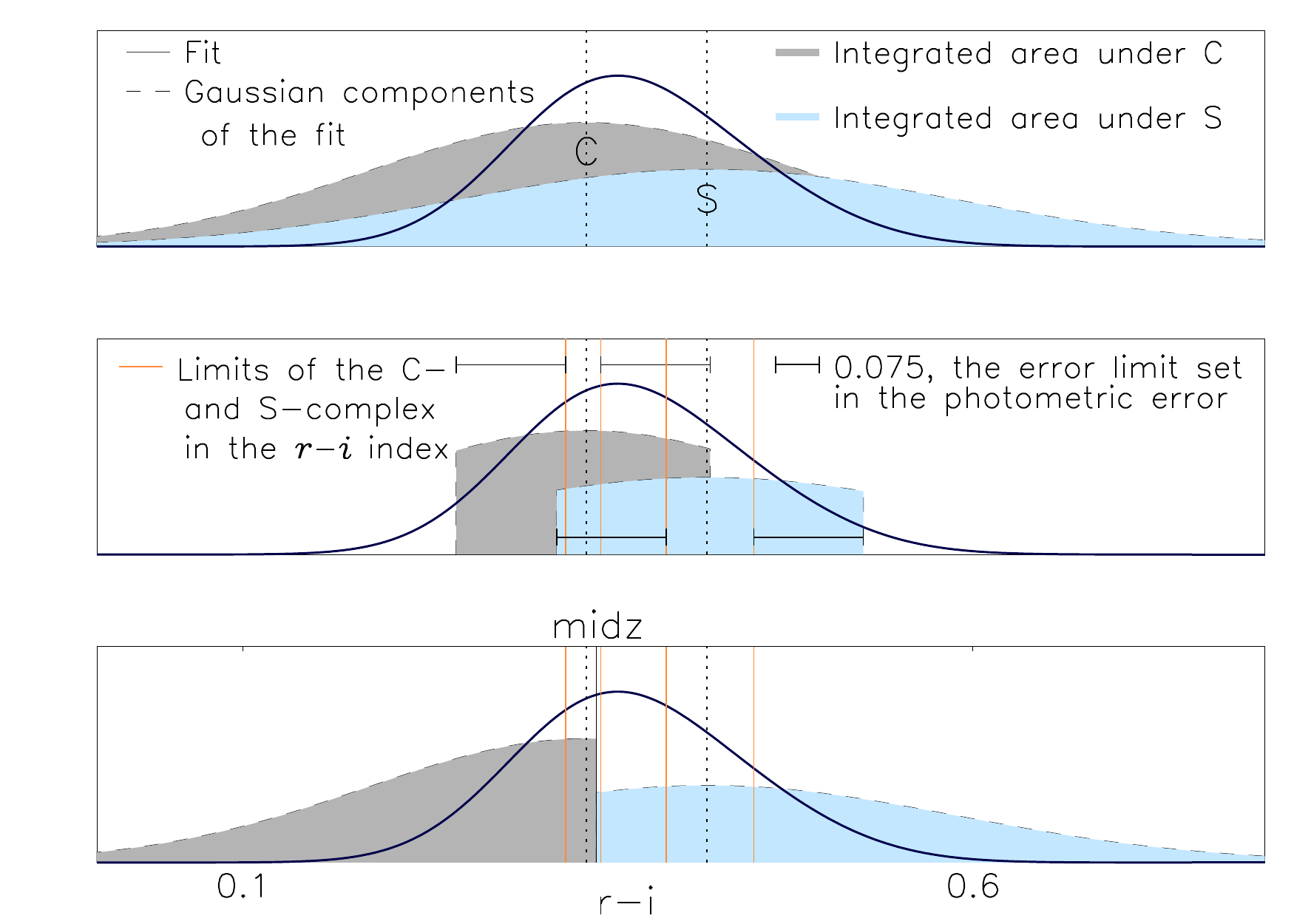}
 \caption{Integration limits considered for the Gaussian components of the fit. The fit is equivalent to the one showed in Figure \ref{fig:rpdf}. Vertical dotted shows the position of the main subtype of the C- and S-complex, thus the center of the Gaussian components. The width of the Gaussians is related to the error of the individual elements, the ordinate axis has no practical meaning.  See text for more details.}
\label{fig:limits}
\end{figure}

As can be seen in Figure \ref{fig:limits}, the width of both the C and S Gaussian distributions from our fit are wider than that of the PDF. This does not affect the result since we are using the ratio of the areas under the fits, but we explore different limits of integration on the abscissa axis for getting each of the areas:

 \begin{enumerate}[label=(\alph*)]
  \item $[-\infty$,$\infty]$
    \item Integrating under the limits of the taxonomic types plus the maximum error allowed in color (0.075).
    \item $[-\infty$,mid$_z]$ for the C Gaussian  and [mid$_z$,$\infty$] for the S Gaussian.
 \end{enumerate}
 
  \noindent where mid$_z$ is the middle point between the \rmi index of the C and S taxonomic complexes in terms of the z-score.\\
 
 The three integration limits are shown in Figure \ref{fig:limits}. They yield a similar $\rm C/S$ ratio, but, method (b) proved to be more stable as a function of the pollution in test runs of the MC\footnote{We used Equations \ref{eq:err} - \ref{eq:sdev} presented in Section \ref{sec:accuracy} to compare the results.} analysis. With the use of this method we made a cut in the tails of the Gaussians, obtaining more localized components. Notice from Figure \ref{fig:types} that the main types of the C- and S-complex are positioned nearly at the center of their corresponding complex range, making the integration reliable. \\

\begin{figure}
\centering
 \includegraphics[clip=true,trim=5 0 10 0,scale=0.7]{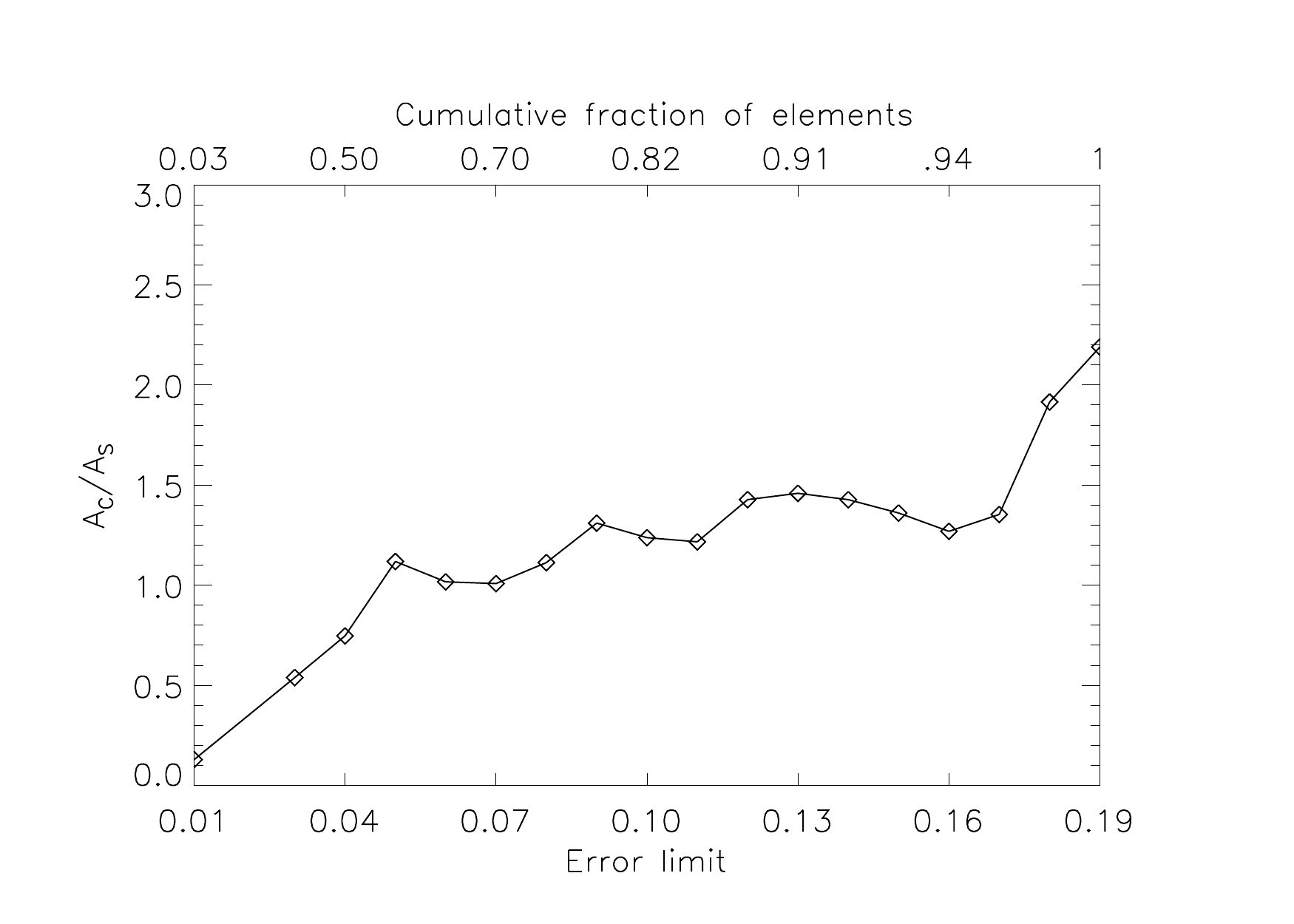}
 \caption{C-type to S-type ratio found with a Gaussian fit to the PDF after using different upper error selection criteria. Notice that this figure is a comparison of different error limits for our sample and it does not represent our main results.}
\label{fig:ratio}
\end{figure}

We explored the $\rm C/S$ ratio (obtained in Section \ref{sec:pdf}) as a function of the upper error allowed in the clean sample. This dependence is shown in Figure \ref{fig:ratio}. Excluding the extremes of the abscissa range on this plot, the $\rm C/S$ ratio does not present a strong dependence on the error limit. Additionally, more than $\mathbf{70\%}$ of the objects that passed our other selection criteria have an error lower than 0.075 (and most of the asteroids in our full sample too, see Figure \ref{fig:errors}).\\

\subsection{Accuracy}\label{sec:accuracy}

The composition of the RPDF is generated during the simulation, therefore it is known per se. Hence, by applying a Gaussian fit to the RPDF, such as the one described in Section \ref{sec:pdf}, we can obtain the reliability of the fit in a particular case with the relative error:

 \begin{equation}\label{eq:err}
e  = \frac{f_{\rm m} -  f_{\rm k}}{f_ {\rm k}},
\end{equation}

\noindent where $f_{\rm k}$ stands for the known compositional fraction of a certain type in the RPDF, and $f_{\rm m}$ for the fraction measured through the Gaussian fit. Then we can consider all of the instances of a fixed k in the $10^7$ runs with:

\begin{equation}\label{eq:eps}
\epsilon  = \overline{e},
\end{equation}

\noindent and the spread of Equation \ref{eq:eps} is measured with:
\begin{equation}\label{eq:sdev}
\sigma  = \sqrt{\overline{e^2} - \overline{e}^2}.
\end{equation}

 Equations \ref{eq:err} -\ref{eq:sdev} are identical for the three complexes and the Q-type. \\

As a general trend, $\epsilon \ $  was larger for the C complex than for the S one. This suggests that if objects
of the X-complex and Q-type are present in the data sample, is more likely for them to be identified
as members of the C-complex than the S-complex by performing a Gaussian fit. This was
also observed in test runs where only the main type of each complex was used. In terms of the \rmi \
index, the Q-type is closer to the C-complex, while the X-complex is closer to the S (see Figure
\ref{fig:types}). It is possible that this behavior is due to the relative width of the C- and S-complexes
and not to the position of the X-complex itself.\\

  The errors reported in Table \ref{tab:results} were obtained by taking the standard deviation of each complex/type
from the set of 12 best matches from the MC simulation described in previous subsection.\\

   The Gaussian fit is not as robust as the MC analysis for obtaining the results, therefore
we can expect lower accuracy. Having a fit centered on the C and S taxonomic components, the ratio
of the area under the corresponding Gaussians is directly related to the compositional fraction of
the sample. The fraction of S-type elements obtained through a Gaussian fit is more reliable than
the C-type one. For this reason, from the fit we focus on the S-type fraction obtained:
$48\%$, which within the error bars is compatible with our MC result.\\

Although in our simulations we allowed for a wide range of variation in the randomly generated
sample, our results partially rely on the assumptions described in Section  \ref{sec:mc}. \\

 \section{Conclusions and future outlook}
 \label{sec:conc}
With the use of spectrophotometry on a 1.5 m robotic telescope, we performed rapid-response observations of small Near Earth Objects. Here we present the results from our optical sample. Measurements were simultaneously made in the \textit{r} and \textit{i} band. After applying selection criteria, our sample
consisted on 131 observations of 82 different NEOs within the size range of $\sim$30-850m.\\

 For the size range considered, we found that C-type asteroids are same as frequent than the S-type, finding $\rm{C/S} = 1.06$. Together, these two asteroid types represent  $\sim 90\%$ of our sample, with the rest likely to be Q- and X-type asteroids. This compositional fraction is in agreement with the results of our previous publications \citep{Mommert,Nic} which are based on UKIRT and KMTNet-SAAO observations.\\

Observations from our program are ongoing. The facility used in this study is now observing with the \textit{Z, Y, J} and \textit{H} near infrared bands in addition to the optical \textit{r} and \textit{i}. By analyzing the data
set presented here, we created the tools to analyze the observations from the rest of the campaign
(2016-ongoing). Future publications from this study will include observations from multiple photometric bands, which will improve the accuracy of the results.\\

\section{Acknowledgments}

 We would like to thank Carlos Román for his help on scheduling RATIR\textquotesingle s observations. 
 We thank the anonymous referee for their valuable comments on this work. SNM wants to dedicate this paper to coauthor, professor and friend B\'arbara Pichardo, RIP.\\
 
 The data used in this paper were totally or partially acquired using the RATIR instrument, funded by the University of California (UC) and NASA Goddard Space Flight Center (GSFC), on the 1.5 meter telescope at Observatorio Astron\'omico Nacional, San Pedro Martir, operated and maintained by OAN-SPM and IA-UNAM. This project was supported in part by the National Aeronautics and Space Administration under the Grant No.NNX15AE90G issued through the SSO Near Earth Object Observations Program. SNM also acknowledges the grant UNAM- DGAPA PAPIIT IN107316. \\
 

\allauthors


\begin{thebibliography}{}
\bibitem[Becerra et al.(2017)]{Becerra} Becerra, R.~L., Watson, A.~M., Lee, W.~H., et al.\ 2017, \apj, 837, 116 
\bibitem[Binzel et al.(2015)]{aiv} Binzel, R.~P., Reddy, V., \& Dunn, T.~L.\ 2015, Asteroids IV, 243 
\bibitem[Binzel et al.(2018)]{Binzel} R.P. Binzel and F.E. DeMeo and S.J. Bus and A. Tokunaga and T.H. Burbine and C. Lantz and D. Polishook and B. Carry and A. Morbidelli and M. Birlan and P. Vernazza and B.J. Burt and N. Moskovitz and S.M. Slivan and C.A. Thomas and A.S. Rivkin and M.D. Hicks and T. Dunn and V. Reddy and J.A. Sanchez and M. Granvik and T. Kohout\ 2018, \icarus
\bibitem[Brown et al.(2013)]{chelya} Brown, P.~G., Assink, J.~D., Astiz, L., et al.\ 2013, \nat, 503, 238 
\bibitem[Butler et al.(2012)]{ratir} Butler, N., Klein, C., Fox, O., et al.\ 2012, \procspie, 8446, 844610 
\bibitem[Butler et al.(2017a)]{nat1} Butler, N., Watson, A.~M., Kutyrev, A., et al.\ 2017, GRB Coordinates Network, Circular Service, No.~21915, \#1 (2017), 21915, 1 
\bibitem[Butler et al.(2017b)]{nat2} Butler, N., Watson, A.~M., Kutyrev, A., et al.\ 2017, GRB Coordinates Network, Circular Service, No.~22061, \#1-2018 (2017), 22061, 1 
\bibitem[Butler et al.(2017c)]{nat3} Butler, N., Watson, A.~M., Kutyrev, A., et al.\ 2017, GRB Coordinates Network, Circular Service, No.~22182, \#1 (2017), 22182, 1 
\bibitem[Carry et al.(2016)]{Carry16} Carry, B., Solano, E., Eggl, S., \& DeMeo, F.~E.\ 2016, \icarus, 268, 340 
\bibitem[Carvano et al.(2010)]{Carvano} Carvano, J.~M., Hasselmann, P.~H., Lazzaro, D., \& Moth{\'e}-Diniz, T.\ 2010, \aap, 510, A43 
\bibitem[DeMeo et al.(2009)]{Demeo} DeMeo, F.~E., Binzel, R.~P., Slivan, S.~M., \& Bus, S.~J.\ 2009, \icarus, 202, 160 
\bibitem[Delsemme(1991)]{Delse} Delsemme, A.~H.\ 1991, IAU Colloq.~116: Comets in the post-Halley era, 167, 377 
\bibitem[Erasmus et al.(2017)]{Nic} Erasmus, N., Mommert, M., Trilling, D.~E., et al.\ 2017, \aj, 154, 162 
\bibitem[Ieva et al.(2018)]{Ieva} Ieva, S., Dotto, E., Epifani, E.~M., et al.\ 2018, \aap, 615, A127 
\bibitem[Galache et al.(2015)]{Galache} Galache, J.~L., Beeson, C.~L., McLeod, K.~K., \& Elvis, M.\ 2015, \planss, 111, 155
\bibitem[Garc{\'{\i}}a-D{\'{\i}}az et al.(2014)]{Tere} Garc{\'{\i}}a-D{\'{\i}}az, M.~T., Gonz{\'a}lez-Buitrago, D., L{\'o}pez, J.~A., et al.\ 2014, \aj, 148, 57 
\bibitem[Giorgini et al.(1997)]{Giorgini} Giorgini, J.~D., Yeomans, D.~K., Chamberlin, A.~B., et al.\ 1997, \baas, 29, 1099 
\bibitem[Hinkle et al.(2015)]{mary} Hinkle, M.~L., Moskovitz, N., Trilling, D., et al.\ 2015, AAS/Division for Planetary Sciences Meeting Abstracts \#47, 47, 301.04 
\bibitem[Ivezi{\'c} et al.(2001)]{Ivezic} Ivezi{\'c}, {\v Z}., Tabachnik, S., Rafikov, R., et al.\ 2001, \aj, 122, 2749 
\bibitem[Littlejohns et al.(2015)]{LJ} Littlejohns, O.~M., Butler, N.~R., Cucchiara, A., et al.\ 2015, \mnras, 449, 2919 
\bibitem[Lin et al.(2018)]{Lin} Lin, C.-H., Ip, W.-H., Lin, Z.-Y., et al.\ 2018, \planss, 152, 116 
\bibitem[Malhotra(1997)]{Malhotra} Malhotra, R.\ 1997, Bulletin of the American Astronomical Society, 29, 23.02 
\bibitem[McAdam et al.(2018)]{Maggie} McAdam, M.~M., Sunshine, J.~M., Howard, K.~T., et al.\ 2018, Lunar and Planetary Science Conference, 49, 2081 
\bibitem[Mommert et al.(2016)]{Mommert} Mommert, M., Trilling, D.~E., Borth, D., et al.\ 2016, \aj, 151, 98 
\bibitem[Moskovitz et al.(2015)]{Moskovitz} Moskovitz, N., Thirouin, A., Binzel, R., et al.\ 2015, IAU General Assembly, 22, 2255616 
\bibitem[Popescu et al.(2018)]{Popescu} Popescu, M., Perna, D., Barucci, M.~A., et al.\ 2018, \mnras, 477, 2786 
\bibitem[Ricci et al.(2015)]{Ricci} Ricci, D., Ram{\'o}n-Fox, F.~G., Ayala-Loera, C., et al.\ 2015, \pasp, 127, 143 
\bibitem[Stuart \& Binzel(2004)]{SandB} Stuart, J.~S., \& Binzel, R.~P.\ 2004, \icarus, 170, 295 
\bibitem[Perna et al.(2018)]{Perna} Perna, D., Barucci, M.~A., Fulchignoni, M., et al.\ 2018, \planss, 157, 82 
\bibitem[Ryan et al.(2015)]{Ryan} Ryan, E.~L., Mizuno, D.~R., Shenoy, S.~S., et al.\ 2015, \aap, 578, A42 
\bibitem[Tapia et al.(2014)]{Tapia} Tapia, M., Rodr{\'{\i}}guez, L.~F., Tovmassian, G., et al.\ 2014, \rmxaa, 50, 127
\bibitem[Taylor (1997)]{Taylor} Taylor, John R. \ 1997, An introduction to error analysis, 2nd Ed, p. 176
\bibitem[Thomas et al.(2011)]{ENV} Thomas, C.~A., Trilling, D.~E., Emery, J.~P., et al.\ 2011, \aj, 142, 85 
\bibitem[Thomas et al.(2014)]{Thomas14} Thomas, C.~A., Emery, J.~P., Trilling, D.~E., et al.\ 2014, \icarus, 228, 217 
\bibitem[Watson et al.(2012)]{auto} Watson, A.~M., Richer, M.~G., Bloom, J.~S., et al.\ 2012, \procspie, 8444, 84445L 
\bibitem[Warner et al.(2009)]{Warner} Warner, B.~D., Harris, A.~W., \& Pravec, P.\ 2009, \icarus, 202, 134 
\end{thebibliography}
\end{document}